# GNSS Outlier Mitigation Via Graduated Non-Convexity Factor Graph Optimization

Weisong Wen, Guohao Zhang and Li-Ta Hsu*

*Abstract*— Accurate and globally referenced global navigation satellite system (GNSS) based vehicular positioning can be achieved in outlier-free open areas. However, the performance of GNSS can be significantly degraded by outlier measurements, such as multipath effects and non-line-of-sight (NLOS) receptions arising from signal reflections of buildings. Inspired by the advantage of batch historical data in resisting outlier measurements, in this paper, we propose a graduated non-convexity factor graph optimization (FGO-GNC) to improve the GNSS positioning performance, where the impact of GNSS outliers is mitigated by estimating the optimal weightings of GNSS measurements. Different from the existing local solutions, the proposed FGO-GNC employs the non-convex Geman McClure (GM) function to globally estimate the weightings of GNSS measurements via a coarse-to-fine relaxation. The effectiveness of the proposed method is verified through several challenging datasets collected in urban canyons of Hong Kong using automobile level and low-cost smartphone level GNSS receivers.

*Index Terms*— GNSS; NLOS; Navigation; Factor graph optimization; Graduated non-convexity; Adaptive tunning; Non-convex Geman McClure; Urban canyons

I. INTRODUCTION

Global navigation satellite system (GNSS) positioning is widely used by systems with navigation requirements, such as the unmanned aerial vehicle [1] and autonomous driving vehicle (ADV) [2, 3]. With the increased availability of multiple satellite constellations, GNSS can provide positioning service with a satisfactory positioning accuracy [4] in the open area. Unfortunately, the positioning accuracy is significantly degraded in highly-urbanized cities such as Hong Kong, due to signal reflection caused by surrounding buildings and even the dynamic objects such as double-decker buses [5]. If the signal direct line-of-sight (LOS) is blocked, and reflected signals from the same satellite are received, the notorious non-line-of-sight (NLOS) receptions occur. The multipath effects are caused when both direct and reflected GNSS signals are received. The NLOS and multipath effects make up the majority of the GNSS outlier measurements, which limit the accuracy of urban GNSS positioning. To solve this problem, numerous methods were studied to mitigate the effects of GNSS outliers. The popular existing solutions can be mainly divided into two categories: (1) GNSS outlier mitigation by sensing the surrounding environment using additional information, such as 3D building models [6], sky-pointing camera [7-9], and 3D LiDAR sensors [10, 11]; (2) GNSS outlier mitigation based on robust models by making use of the historical batch data, such as the switchable constraints [12], and dynamic covariance estimation [13]. The following sub-section presents the existing works that lie in those two categories.

*A. Additional Information Aided GNSS Outlier Mitigation*

Since the GNSS outlier measurements are mainly caused by the surrounding buildings, researchers firstly proposed to utilize 3D building models to detect NLOS signals and then exclude the detected NLOS signals for use in GNSS positioning [14-16]. However, NLOS exclusion could distort the perceived geometric distribution of the satellites, degrading accuracy [17], and even resulting in an insufficient number of satellites for further GNSS calculation [17]. Moreover, these methods require the availability of 3D building models of the environment, and the performance of NLOS detection relies on the accuracy of an initial guess of the GNSS receiver's position. Besides, NLOS reception caused by dynamic objects cannot be detected as well. The continuously improved 3D mapping information-aided GNSS positioning methods are studied, like GNSS shadow matching[18-20], likelihood-based method [21], a combination of GNSS shadow matching, and likelihood-based method [21], range-based 3DMA GNSS [22-26]. However, all these solutions require prior 3D mapping information and are sensitive to the initial guess of the GNSS receiver's position. A detailed review of the recent progress for the first category can be found in our previous work in [11].

*B. Robust Model Aided GNSS Outlier Mitigation*

Instead of using additional 3D building models, a team from the Chemnitz University of Technology employed the robust model [13, 27, 28] to mitigate the effects of outlier measurements in GNSS positioning. They opened a new insight to detect the GNSS outlier measurements by taking advantage of the historical GNSS measurements. In their earliest work, as reported in [29], a state-of-the-art factor graph optimization (FGO) [30], which makes use of all of the available historical GNSS measurements to estimate the state set of the GNSS receiver, was applied to solve the GNSS positioning problem. The improved performance was obtained compared with the conventional weighted least squares (WLS) approach. Interestingly, the work described in [31] included a switchable constraint in the factor graph to model the probability of one satellite being an outlier measurement, either multipath or NLOS receptions. In [13], the dynamic covariance estimation (DCE) is proposed where the covariance of GNSS

Weisong Wen and Li-Ta Hsu, are with Hong Kong Polytechnic University, Hong Kong (correspondence e-mail: lt.hsu@polyu.edu.hk).

measurements is treated as an unknown variable to be estimated in the FGO. As a result, the outlier GNSS measurements are de-weighted from the FGO. Considering the non-Gaussian property of the GNSS pseudorange measurements, a self-tuning Gaussian mixture model (GMM) [27, 28] has been proposed to model the uncertainty of GNSS pseudorange measurements. Improved accuracy is obtained compared with switchable constraints, or DCE. Unfortunately, both the switchable constraint [31], DCE [13], and self-tuning GMM only derive local solutions that rely heavily on the initial guess of the state estimation. Moreover, the GMM introduces additional non-convexity for the FGO, making it getting into the local minimum easily. In short, these robust models aided GNSS outlier mitigation methods explored to make use of the historical data to facilitate the outlier detection. However, without the reliance on an accurate initial guess [32], a local optimal solution could be obtained.

Recently, the team from the Massachusetts Institute of Technology proposed a graduated non-convexity (GNC) aided robust and global outlier rejection for point cloud registration [32]. The point cloud registration is solved by formulating the robust least-square estimation as the combination of weighted least squares and the outlier process using the Black-Rangarajan Duality [33]. The work solves the non-convexity issue arising from the Geman McClure function using the GNC and enables the global and optimal estimation of the weightings of corresponding measurements simultaneously. However, the proposed method in [32] only considered the measurements in a single epoch. In other words, the advantage of historical information is not fully explored. Moreover, there exists a distinct boundary between the inlier and outlier measurements in the evaluated dataset which limits the challenges for detecting the outlier measurements, and its potential in other field is still to be explored.

Inspired by the innovative work in [32], this paper goes one step further by extending the work in [32] to GNSS pseudorange and Doppler measurements integration using FGO. Different from the problem evaluated in [32], the boundary between the inlier and outlier GNSS measurements is ambiguous and highly environmental dependant. Moreover, the historical information is considered simultaneously to explore the time-correlation among GNSS measurements.

The main contributions of this paper are summarized as follows:

(1) We derive the combined formulation (FGO-GNC) of (a) FGO-based GNSS pseudorange and Doppler measurements integration and (b) the adaptive weighting estimation (AWE), where the AWE is achieved based on the non-convex Geman McClure (GM) function aided by GNC.

(2) We evaluated the proposed method using several challenging datasets collected in urban canyons of Hong Kong. Meanwhile, we present the details of how the GNC helps to get rid of the local minimum via a coarse-to-fine process and globally estimate the optimal weightings.

(3) Both commercial and smartphone level GNSS receivers are used in the experiments. It shows the potential of the proposed GNSS positioning algorithms in mass-market applications.

To the best of the authors' knowledge, this paper is the first work that presents the GNC-aided FGO formulation to globally estimate the state and optimal weighting of GNSS measurements simultaneously.

The remainder of this paper is organized as follows. An overview of the proposed method is given in Section II. For completeness, we present the GNSS pseudorange and Doppler measurements fusion using FGO in Section III. In Section IV, the detail of the proposed GNC-aided FGO formulation with the weighting estimation is presented. Several real experiments were performed to evaluate the effectiveness of the proposed method in Section V. Finally, conclusions are drawn, and further work is presented in Section VI.

## II. OVERVIEW OF THE PROPOSED METHOD

An overview of the method proposed in this paper is shown in Fig. 1. The inputs of the system are the pseudorange and Doppler frequency measurements from the GNSS receiver. The output is the state estimation of the GNSS receiver. The system mainly consists of two parts, the FGO, and the AWE. The FGO integrates the GNSS pseudorange and the Doppler measurements aided by the weighting from the AWE. The AWE is fed by the residuals from the FGO and outputs the weightings of all the measurements. The FGO and the AWE are performed iteratively until the convergence metric of AWE is satisfied and the whole process is called FGO-GNC in this paper. The FGO for pseudorange and Doppler fusion is detailed in Section III and the FGO-GNC is presented in Section IV.

Matrices are denoted as uppercase with bold letters. Vectors are denoted as lowercase with bold letters. Variable scalars are denoted as lowercase italic letters. Constant scalars are denoted as lowercase letters. To make the proposed pipeline clear, the following major notations are defined and followed by the rest of the paper. Be noted that the state of the GNSS receiver and the position of satellites are all expressed in the earth-centered, earth-fixed (ECEF) frame.

a) The pseudorange measurement received from a satellite $s$ at a given epoch $t$ is expressed as $\rho_{r,t}^s$. The subscript $r$ denotes the GNSS receiver. The superscript $s$ denotes the index of the satellite.
b) The Doppler measurement received from satellite $s$ at a given epoch $t$ is expressed as $d_{r,t}^s$.
c) The position of the satellite $s$ at a given epoch $t$ is expressed as $\mathbf{p}_t^s = (p_{t,x}^s, p_{t,y}^s, p_{t,z}^s)^T$.
d) The velocity of the satellite $s$ at a given epoch $t$ is expressed as $\mathbf{v}_t^s = (v_{t,x}^s, v_{t,y}^s, v_{t,z}^s)^T$.
e) The position of the GNSS receiver at a given epoch $t$ is expressed as $\mathbf{p}_{r,t} = (p_{r,t,x}, p_{r,t,y}, p_{r,t,z})^T$.
f) The velocity of the GNSS receiver at a given epoch $t$ is expressed as $\mathbf{v}_{r,t} = (v_{r,t,x}, v_{r,t,y}, v_{r,t,z})^T$.
g) The clock bias of the GNSS receiver at a given epoch $t$ is expressed as $\delta_{r,t}$, that with the unit in meters. $\delta_{r,t}^s$ denotes the satellite clock bias by meters.

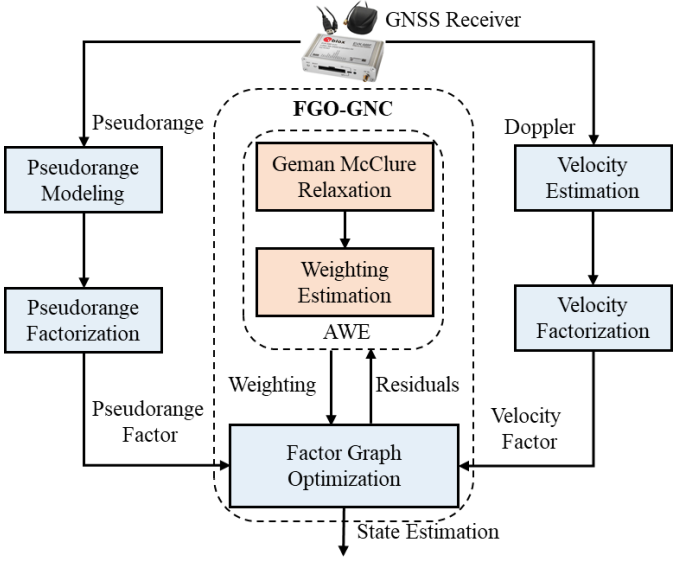

Fig. 1. Overview of the proposed GNSS pseudorange/Doppler integration via FGO-GNC. The light blue area denotes the factor graph optimization part. The orange area represents the adaptive weighting estimation part.

## III. GNSS Positioning Via FGO

The state of the GNSS receiver is represented as follows:

$$\chi = [x_{r,1}, x_{r,2}, ..., x_{r,n}] \quad (1)$$

where the variable $\chi$ denotes the set states of the GNSS receiver ranging from the first epoch to the current $n$. The state of the GNSS receiver at a single epoch can be denoted as follows:

$$x_{r,t} = (p_{r,t}, v_{r,t}, \delta_{r,t})^T \quad (2)$$

where the $x_{r,t}$ denotes the state of the GNSS receiver at epoch $t$ which involves the position ($p_{r,t}$), velocity ($v_{r,t}$) and receiver clock bias ($\delta_{r,t}$).

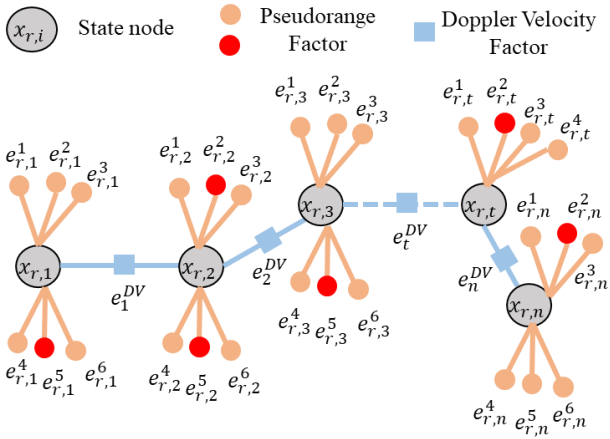

Fig. 2. The orange (healthy) and red (outlier) circles denote the pseudorange factor (e.g $e_{r,t}^s$). The light blue shaded rectangle represents the Doppler velocity factor (e.g $e_t^{DV}$). The grey shaded circle stands for the state of the GNSS receiver.

The graph structure of the proposed FGO for solving the GNSS pseudorange/Doppler integration is shown in Fig. 2. The subscript $n$ denotes the total epochs of measurements considered in the FGO. Each state in the factor graph is connected using the Doppler velocity factor. Note that the formulation of the GNSS pseudorange/Doppler integration using the FGO is firstly presented in our latest work in [34]. However, we still present it here for the completeness of the derivation of FGO-GNC in Section IV.

### A. Pseudorange Measurement Modeling

The pseudorange measurement from the GNSS receiver, $\rho_{r,t}^s$, is denoted as follows [35].

$$\rho_{r,t}^s = r_{r,t}^s + c(\delta_{r,t} - \delta_{r,t}^s) + I_{r,t}^s + T_{r,t}^s + \varepsilon_{r,t}^s \quad (3)$$

where $r_{r,t}^s$ is the geometric range between the satellite and the GNSS receiver. $I_{r,t}^s$ represents the ionospheric delay distance; $T_{r,t}^s$ indicates the tropospheric delay distance. $\varepsilon_{r,t}^s$ represents the errors caused by the multipath effects, NLOS receptions, receiver noise, antenna phase-related noise. Meanwhile, the atmosphere effects ($T_{r,t}^s$ and $I_{r,t}^s$) are compensated using the conventional models (Saastamoinen and Klobuchar models, respectively) presented in RTKLIB [36].

The observation model for GNSS pseudorange measurement from a given satellite $s$ is represented as follows:

$$\rho_{r,t}^s = h_{r,t}^s(p_{r,t}, p_t^s, \delta_{r,t}) + \omega_{r,t}^s \quad (4)$$

with $h_{r,t}^s(p_{r,t}, p_t^s, \delta_{r,t}) = ||p_t^s - p_{r,t}|| + \delta_{r,t}$

where the variable $\omega_{r,t}^s$ stands for the noise associated with the $\rho_{r,t}^s$. Therefore, we can get the error function ($e_{r,t}^s$) for a given satellite measurement $\rho_{r,t}^s$ as follows:

$$||e_{r,t}^s||_{\Sigma_{r,t}^s}^2 = ||\rho_{r,t}^s - h_{r,t}^s(p_{r,t}, p_t^s, \delta_{r,t})||_{\sigma_{r,t}^s}^2 \quad (5)$$

where $\sigma_{r,t}^s$ denotes the covariance. We calculate the $\sigma_{r,t}^s$ based on the satellite elevation angle, signal, and noise ratio (SNR) following the work in [37].

### B. Doppler Measurements Modeling

Given the Doppler measurement ($d_{r,t}^1, d_{r,t}^2, ...$) of each satellite at an epoch $t$, the velocity ($v_{r,t}$) of the GNSS receiver can be calculated using the WLS method [38]. Giving that the state of the velocity, $x_t^d$, is as follows:

$$x_t^d = (v_{r,t}, \dot{\delta}_{r,t})^T \quad (6)$$

where the $v_{r,t}$ represents the velocity of the GNSS receiver. The variable $\dot{\delta}_{r,t}$ stands for the receiver clock drift. The range rate measurement vector ($y_{r,t}^d$) at an epoch $t$ is expressed as follows:

$$y_{r,t}^d = (\lambda d_{r,t}^1, \lambda d_{r,t}^2, \lambda d_{r,t}^3, ...)^T \quad (7)$$

where the $\lambda$ denotes the carrier wavelength of the satellite signal, the $d_{r,t}^s$ represents the Doppler measurement. The observation function $h^d(*)$ which connects the state, and the Doppler measurements are expressed as follows:

$$h^d(x_t^d) = \begin{bmatrix} rr_{r,t}^1 + \dot{\delta}_{r,t} - \dot{\delta}_{r,t}^1 \\ rr_{r,t}^2 + \dot{\delta}_{r,t} - \dot{\delta}_{r,t}^2 \\ rr_{r,t}^3 + \dot{\delta}_{r,t} - \dot{\delta}_{r,t}^3 \\ ... \\ rr_{r,t}^m + \dot{\delta}_{r,t} - \dot{\delta}_{r,t}^m \end{bmatrix} \quad (8)$$

where the superscript $m$ denotes the total number of satellites. The variable $rr_{r,t}^m$ denotes the expected range rate. The variables $\dot{\delta}_{r,t}$ and $\dot{\delta}_{r,t}^m$ represent the receiver and satellite clock bias drift. The expected range rate $rr_{r,t}^s$ for satellite $s$ can also be calculated as follows:

$$rr_{r,t}^s = \mathbf{e}_{r,t}^{s,LOS}(\mathbf{v}_t^s - \mathbf{v}_{r,t}) + \frac{\omega_{earth}}{c_L}(v_{t,y}^s p_{r,t,x} + p_{t,y}^s v_{r,t,x} - p_{t,x}^s v_{r,t,y} - v_{t,x}^s p_{r,t,y}) \quad (9)$$

where the variable $\omega_{earth}$ denotes the angular velocity of the earth's rotation [36]. The variable $c_L$ denotes the speed of light. The variable $\mathbf{e}_{r,t}^{s,LOS}$ denotes the line-of-sight vector connecting the GNSS receiver and the satellite. The Jacobian matrix $\mathbf{H}_t^d$ for the observation function $h^d(*)$ is denoted as follows:

$$\mathbf{H}_t^d = \begin{bmatrix} \frac{p_{t,x}^1-p_{r,t,x}}{\|\mathbf{p}_t^1-\mathbf{p}_{r,t}\|} & \frac{p_{t,y}^1-p_{r,t,y}}{\|\mathbf{p}_t^1-\mathbf{p}_{r,t}\|} & \frac{p_{t,z}^1-p_{r,t,z}}{\|\mathbf{p}_t^1-\mathbf{p}_{r,t}\|} & 1 \\ \frac{p_{t,x}^2-p_{r,t,x}}{\|\mathbf{p}_t^2-\mathbf{p}_{r,t}\|} & \frac{p_{t,y}^2-p_{r,t,y}}{\|\mathbf{p}_t^2-\mathbf{p}_{r,t}\|} & \frac{p_{t,z}^2-p_{r,t,z}}{\|\mathbf{p}_t^2-\mathbf{p}_{r,t}\|} & 1 \\ \frac{p_{t,x}^3-p_{r,t,x}}{\|\mathbf{p}_t^3-\mathbf{p}_{r,t}\|} & \frac{p_{t,y}^3-p_{r,t,y}}{\|\mathbf{p}_t^3-\mathbf{p}_{r,t}\|} & \frac{p_{t,z}^3-p_{r,t,z}}{\|\mathbf{p}_t^3-\mathbf{p}_{r,t}\|} & 1 \\ \dots & \dots & \dots & \dots \\ \frac{p_{t,x}^m-p_{r,t,x}}{\|\mathbf{p}_t^m-\mathbf{p}_{r,t}\|} & \frac{p_{t,y}^m-p_{r,t,y}}{\|\mathbf{p}_t^m-\mathbf{p}_{r,t}\|} & \frac{p_{t,z}^m-p_{r,t,z}}{\|\mathbf{p}_t^m-\mathbf{p}_{r,t}\|} & 1 \end{bmatrix} \quad (10)$$

where the operator $\|*\|$ is employed to calculate the range distance between the given satellite and the GNSS receiver. Therefore, the velocity ($\mathbf{v}_{r,t}$) of the GNSS receiver can be estimated via least squares iteratively based on (8), (9), and (10) as follows:

$$\mathbf{x}_t^d = (\mathbf{H}_t^{d^T}\mathbf{H}_t^d)^{-1}\mathbf{H}_t^{d^T}\mathbf{y}_{r,t}^d \quad (11)$$

The observation model for the velocity ($\mathbf{v}_{r,t}$) is expressed as follows:

$$\mathbf{v}_{r,t}^{DV} = h_{r,t}^{DV}(\mathbf{x}_{r,t+1}, \mathbf{x}_{r,t}) + \omega_{r,t}^{DV} \quad (12)$$

with $h_{r,t}^{DV}(\mathbf{x}_{r,t+1}, \mathbf{x}_{r,t}) = \begin{bmatrix} (p_{r,t+1,x} - p_{r,t,x})/\Delta t \\ (p_{r,t+1,y} - p_{r,t,y})/\Delta t \\ (p_{r,t+1,z} - p_{r,t,z})/\Delta t \end{bmatrix}$

where the $\mathbf{v}_{r,t}^{DV}$ denotes the velocity measurements given by the estimation in (11). The variable $\omega_{r,t}^{DV}$ denotes the noise associated with the velocity measurement. The variable $\Delta t$ denotes the time difference between epoch $t$ and epoch $t + 1$. Therefore, we can get the error function ($\mathbf{e}_{r,t}^{DV}$) for a given Doppler velocity measurement $\mathbf{v}_{r,t}^{DV}$ as follows:

$$\|\mathbf{e}_{r,t}^{DV}\|_{\Sigma_{r,t}^{DV}}^2 = \|\mathbf{v}_{r,t}^{DV} - h_{r,t}^{DV}(\mathbf{x}_{r,t+1}, \mathbf{x}_{r,t})\|_{\Sigma_{r,t}^{DV}}^2 \quad (13)$$

where $\mathbf{\Sigma}_{r,t}^{DV}$ denotes the covariance matrix corresponding to the Doppler velocity measurement.

### C. Pseudorange/Doppler Fusion Via FGO

In this paper, we assume that the pseudorange and the Doppler measurements are independent of each other for different satellites [39]. Based on the factors derived above, the objective function for the GNSS pseudorange/Doppler integration using FGO is formulated as follows:

$$\boldsymbol{\chi}^* = \arg\min_{\boldsymbol{\chi}} \sum_{s,t}(\|\mathbf{e}_{r,t}^{DV}\|_{\Sigma_{r,t}^{DV}}^2 + \|e_{r,t}^s\|_{\sigma_{r,t}^s}^2) \quad (14)$$

The variable $\boldsymbol{\chi}^*$ denotes the optimal estimation of the state sets, which can be estimated by solving the objective function above. Essentially, the problem of (14) is a non-linear least-square problem that is sensitive to the gross outlier measurements. Fig. 3 shows a histogram of the pseudorange residual after solving the (14) based on a dataset collected in an urban canyon of Hong Kong. We can see that the histogram involves a long tail of the left side, which is mainly caused by the GNSS outlier measurements. In the next section, we present the FGO-GNC to mitigate the effects of the GNSS outlier measurements.

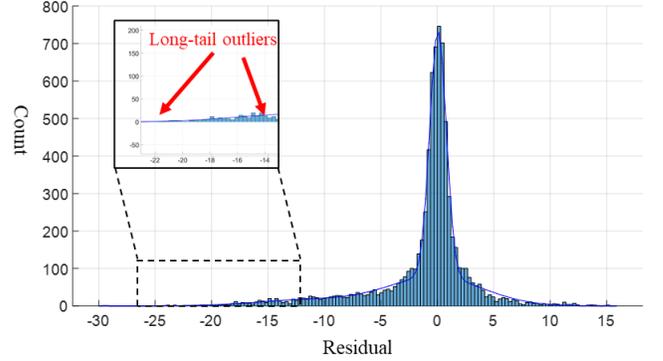

Fig. 3. Histogram of the pseudorange residuals after solving (14) using FGO based on a dataset collected in an urban canyon of Hong Kong. The x-axis denotes the residual values. The y-axis denotes the number of counts.

## IV. OUTLIER-AWARE GNSS PSEUDORANGE/DOPPLER INTEGRATION

### A. Formulation

As the Doppler measurements are less sensitive to the multipath and NLOS receptions, we mainly consider the outliers in the pseudorange measurements. Referring to (14), the robustified objective function can be formulated as follows.

$$\boldsymbol{\chi}^* = \arg\min_{\boldsymbol{\chi}} \left( \sum_{s,t}\|\mathbf{e}_{r,t}^{DV}\|_{\Sigma_{r,t}^{DV}}^2 + \Psi(\|e_{r,t}^s\|_{\sigma_{r,t}^s}) \right) \quad (15)$$

where the function $\Psi(*)$ is a robust function. According to the Black-Rangarajan Duality [33], a robustified non-linear least square problem (15) is equivalent to the following decoupled formulation:

$$\boldsymbol{\chi}^* = \arg\min_{\boldsymbol{\chi},\boldsymbol{\mathcal{W}}} \sum_{s,t}\left(\|\mathbf{e}_{r,t}^{DV}\|_{\Sigma_{r,t}^{DV}}^2 + \omega_{t,s}\|e_{r,t}^s\|_{\sigma_{r,t}^s}^2 + \emptyset_\rho(\omega_{t,s})\right) \quad (16)$$

where the $\omega_{t,s}$ denotes the weighting for a given pseudorange measurement from satellite $s$ at the epoch $t$, satisfying $\omega_{t,s} \in [0,1]$. The variable $\boldsymbol{\mathcal{W}}$ is a set of weightings of $\omega_{t,s}$. The function $\emptyset_\rho(\omega_{t,s})$ is the outlier process that encodes the penalty on the weighing $\omega_{t,s}$, which is determined by the chosen robust function. Therefore, the unknowns of the system involve both the $\mathbf{x}_{r,t}$ and the optimal weighting ($\omega_{t,s}$) of the pseudorange measurements. Solving the (16) equals finding the optimal state estimation of the GNSS receiver and the optimal weightings of pseudorange measurements to minimize the summation of the residuals. To simplify the derivation in the rest of this paper, we represent the weighted residual $||\mathbf{e}_{r,t}^s||_{\Sigma_{r,t}^s}$ using $\tilde{e}_{r,t}^s$.

Typically, the robustified loss using the Geman McClure function [40] for the given error function $\tilde{e}_{r,t}^s$ corresponding to the pseudorange measurement from satellite $s$ can be formulated as follows:

$$\Psi(\tilde{e}_{r,t}^s) = \frac{(c_{GM})^2 (\tilde{e}_{r,t}^s)^2}{(c_{GM})^2 + (\tilde{e}_{r,t}^s)^2} \quad (17)$$

where the $c_{GM}$ is the parameter that determines the shape of the Geman McClure function. Fig. 4 shows the Geman McClure loss corresponding to residual ($\tilde{e}_{r,t}^s$) ranging from (-10, 10) with different $c_{GM}$. The smaller $c_{GM}$ introduces stronger resistance against the outliers as the impacts from the enormous outliers are mitigated by the low curvature long tail. However, this can lead to a highly non-convex surface. As a result, it is hard to globally solve the (16) using typical nonlinear least square estimation. To fill this gap, we formulate the GNC-aided FGO to solve the (16) in a coarse-to-fine manner in the next section.

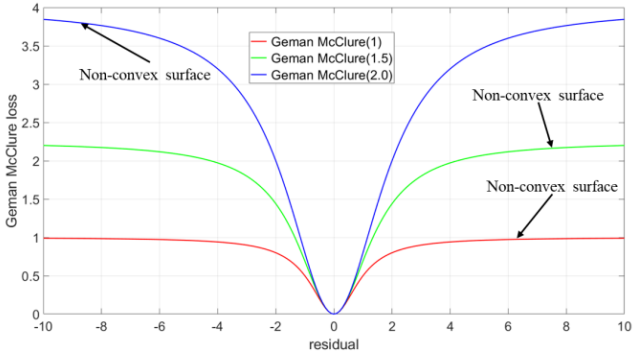

Fig. 4. Illustration of the Geman McClure function with different parameters $c_{GM}$ annotated with different colors (red: $c_{GM} = 1$, green: $c_{GM} = 1.5$, blue: $c_{GM} = 2.0$).

*B. Coarse-to-Fine Solution to FGO-GNC*

One of the solutions to avoid the local minimum arising from the non-convex surface is to directly provide an accurate initial guess. Unfortunately, this is usually hard to be satisfied in the practice. The GNC is a state-of-the-art approach to solve the problem with the non-convex surface. Interestingly, the GNC is firstly used in computer vision-related research [41] and regression problems [42]. The theory is less known in the navigation field. The principle of GNC for solving non-convex optimization is to adopt a set of surrogate functions ($\rho_\theta(\tilde{e}_{r,t}^s)$) to approximate the original non-convex surface. The surrogate function corresponding to (17) is formulated as follows [32]:

$$\rho_\theta(\tilde{e}_{r,t}^s) = \frac{\theta \cdot (c_{GM})^2 (\tilde{e}_{r,t}^s)^2}{\theta \cdot (c_{GM})^2 + (\tilde{e}_{r,t}^s)^2} \quad (18)$$

where the $\theta$ is the control parameter that governing the non-convexity of the $\rho_\theta(\tilde{e}_{r,t}^s)$. When the $\theta$ is large, the $\rho_\theta(\tilde{e}_{r,t}^s)$ has an almost convex surface which is easy to optimize. When the $\theta$ is small (e.g. $\theta = 1$), the $\rho_\theta(\tilde{e}_{r,t}^s)$ recover to the original Geman McClure function as (17) which has a non-convex surface. Therefore, the GNC estimates the solution by starting from its convex surface (with large $\theta$) and gradually relax the $\rho_\theta(\tilde{e}_{r,t}^s)$ until the original non-convex surface is recovered. Meanwhile, the solution estimated during each iteration is employed as the initial guess of the next iteration, leading to a coarse-to-fine optimization.

Fig. 5 shows the surrogate function for Geman McClure with different parameters $\theta$. The non-convexity is recovered gradually with decreased $\theta$. Therefore, (16) can be rewritten as follows:

$$\boldsymbol{\chi}^* = \underset{\boldsymbol{\chi}, \boldsymbol{W}}{\arg\min} \sum_{s,t} \left( ||\mathbf{e}_{r,t}^{DV}||_{\Sigma_{r,t}^{DV}}^2 + \omega_{t,s} ||\mathbf{e}_{r,t}^s||_{\Sigma_{r,t}^s}^2 + \emptyset_{\rho_\theta}(\omega_{t,s}) \right) \quad (19)$$

where the $\emptyset_{\rho_\theta}(\omega_{t,s})$ is the surrogated penalty function. The work in [33] details the calculation of the penalty function based on the selected robust function (e.g. Geman McClure). We select the penalty function as follows [32]:

$$\emptyset_{\rho_\theta}(\omega_{t,s}) = \theta\,(c_{GM})^2 (\sqrt{\omega_{t,s}} - 1)^2 \quad (20)$$

Substituting (20) back to (19) and we can find that the optimal weighting can be solved by calculating the derivative of the objective function (19) concerning the $\omega_{t,s}$ as $G_{t,s}(\omega_{t,s})$:

$$G_{t,s}(\omega_{t,s}) = ||\mathbf{e}_{r,t}^s||_{\Sigma_{r,t}^s}^2 + \theta \cdot (c_{GM})^2 (1 - \frac{1}{\sqrt{\omega_{t,s}}}) \quad (21)$$

Fortunately, the unique solution to $\omega_{t,s}$ can directly be solved from (21) in closed form as follows:

$$\omega_{t,s} = \frac{\theta \cdot (c_{GM})^2}{\theta \cdot (c_{GM})^2 + (\tilde{e}_{r,t}^s)^2} \quad (22)$$

Therefore, the optimal weighting corresponding to a given measurement can be calculated by (22) concerning each surrogate function $\rho_\theta(\tilde{e}_{r,t}^s)$. The detail of solving (19) is presented in Algorithm 1. The input of Algorithm 1 is a set of GNSS measurements $\rho_{r,t}^s$ and $d_{r,t}^s$. The outputs are the state set $\boldsymbol{\chi}$ and optimal weightings $\boldsymbol{W}$. In Step 1, the state is initialized using the WLS. The weightings are initialized by setting to 1. Then (19) is solved by alternating minimization via Steps 1 and 2.

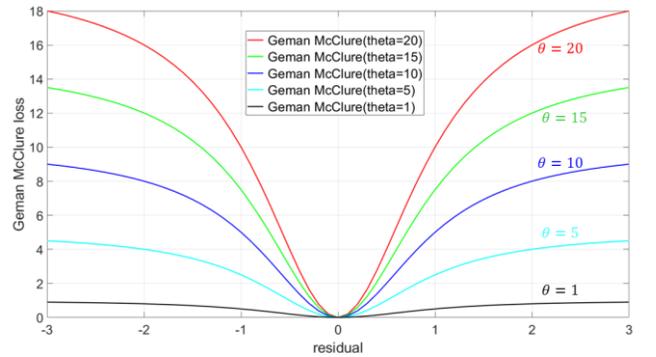

Fig. 5. Illustration of the surrogate function for Geman McClure with different control parameters $\theta$ annotated with different colors (red: $\theta = 20$, green: $\theta = 15$, blue: $\theta = 10$, cyan: $\theta = 5$, black: $\theta = 1$).

***Selection of the initial value for the control parameter $\theta$***: The initial value of the control parameter $\theta$ is significant for the iteration of the FGO-GNC. A too-large initial value of the $\theta$ can lead to unnecessary iterations during the optimization of (19). On the contrary, a too-small initial value of the $\theta$ cannot guarantee that all the residuals are originally located inside the convex surface of the Geman McClure function. Therefore, we propose to select the original value of $\theta$ as (23), based on the maximum value of the pseudorange residual to make sure that all the measurement residuals are located inside the convex area of the Geman McClure function at the first iteration in the Algorithm 1.

$$\theta = \frac{3\,(e_{r,t_{max}}^s)^2}{(c_{GM})^2} \quad (23)$$

where the $e_{r,t_{max}}^s$ denotes the maximum residual among the pseudorange measurements after the initialization of Step 1 in Algorithm 1.

---

**Algorithm 1: Solution to FGO-GNC**

**Inputs**: A set of GNSS measurements $\rho_{r,t}^s$ and $d_{r,t}^s$
**Outputs**: A set of state $\chi$ and optimal weightings $\mathcal{W}$
**Step 1 (Initialization)**: Initialize the state set $\chi$ using the weighted least squares. Initialize the $\mathcal{W}$ by setting all the weighting to 1.
**Step 2 (FGO)**: Solve the following equation to estimate the state $\chi$ with fixed weighting set $\chi$ as follows:

$$\chi^* = \arg\min_{\chi} \sum_{s,t} ||\mathbf{e}_{r,t}^{DV}||_{\Sigma_{r,t}^{DV}}^2 + \omega_{t,s}||e_{r,t}^s||_{\sigma_{r,t}^s}^2 \quad (24)$$

**Step 3 (AWE)**: Update the optimal weighting set ($\chi$) concerning each pseudorange measurement based on (21).
**Step 4**: $\theta = \theta/1.4$; repeat Step 2~3, until $\theta < 1$.

---

### C. Global optimality Analysis of the GNC-FGO Problem

The original GNSS pseudorange/Doppler fusion problem formulated by equation (14) consists of two kinds of constraints (a) the geometry distance constraint from the pseudorange measurement connecting the GNSS receiver and the satellite. (b) the velocity constraint derived from the Doppler measurements which are linear with the states of the GNSS receiver. Meanwhile, the cost function from the geometry distance constraint is locally convex as long as the GNSS receiver is located on the earth's surface [43].

***Global optimality of state estimation***: After applying the robustified function leading to (19), the additional non-convexity is introduced due to the non-convex surface of the Geman McClure function. With the help of the GNC, the initial guess of the state estimation is refined gradually by relaxing the Geman McClure function. Therefore, the global optimization of state estimation is guaranteed [32].

***Global optimality of weighting estimation***: Given the derivative of the objective function concerning the $\omega_{t,s}$ as $G_{t,s}(\omega_{t,s})$, the function $G_{t,s}(\omega_{t,s})$ is a monotonic function. The $G_{t,s}(\omega_{t,s})$ equals to $||\mathbf{e}_{r,t}^s||_{\sigma_{r,t}^s}^2$ when the $\omega_{t,s}$ equals to 1. Meanwhile, the $G_{t,s}(\omega_{t,s})$ equals to $-\infty$ when the $\omega_{t,s} \to 0$. Given all these properties, the unique optimal estimation of the weighting $\omega_{t,s}$ is guaranteed.

## V. EXPERIMENT RESULTS AND DISCUSSION

### A. Experiment Setup

To evaluate the performance of the proposed method in mitigating the effects of GNSS outlier measurements, two experiments were conducted in typical urban canyons in Hong Kong. The data collection vehicle is shown in the following Fig. 6. In both experiments, a commercial u-blox M8T GNSS receiver was used to collect raw single-frequency GPS/BeiDou measurements at a frequency of 1 Hz. The lengths for data collection in the two experiments are 482 (Section V-B) seconds and 1204 seconds (Section V-C), respectively. Besides, the NovAtel SPAN-CPT, a GNSS (GPS, GLONASS, and BeiDou) RTK/INS (fiber-optic gyroscopes, FOG) integrated navigation system was used to provide ground truth of positioning. The gyro bias in-run stability of the FOG is 1 degree per hour, and its random walk is 0.067 degrees per hour. The baseline between the rover and the GNSS base station is within 7 km. All the data were collected and synchronized using a robot operation system (ROS) [44]. The coordinate systems between the GNSS receiver and the NovAtel SPAN-CPT were calibrated before the experiments. All the data are post-processed using a desktop (Intel Core i7-9700K CPU, 3.60 GHz) computer.

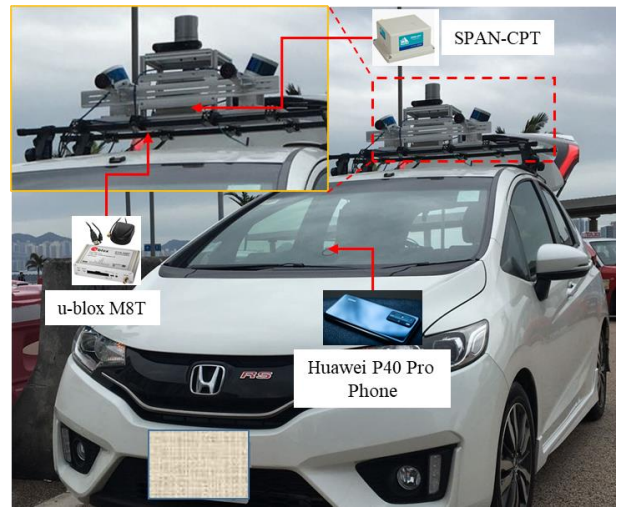

Fig. 6. Data collection vehicle and the sensor setup.

To verify the effectiveness of the proposed method, six positioning schemes were compared and details were presented:
(1) **WLS**: GNSS WLS positioning based on pseudorange [34].
(2) **EKF**: GNSS pseudorange/Doppler integration using EKF [34].
(3) **FGO**: GNSS pseudorange/Doppler integration using FGO [45].
(4) **FGO-Cauchy**: GNSS pseudorange/Doppler integration using FGO aided by non-convex Cauchy loss function [46].
(5) **FGO-GM**: GNSS pseudorange/Doppler integration using FGO aided by non-convex Geman McClure function [46] without using the graduated non-convexity.
(6) **FGO-GNC** (proposed integration): GNSS pseudorange/Doppler integration using FGO aided by FGO-GNC.

This paper firstly presents the GNSS positioning from WLS to show the overall performance. Then the integration of pseudorange and Doppler measurements using filtering (EKF) and FGO together with its improvements are evaluated. Besides the performance evaluation, we also deeply analyze the coarse-to-fine process during the solving of FGO-GNC to further show how the proposed method mitigates the effects of the GNSS outlier measurements.

*Applied parameters*: The shape parameters, also called kernel width, for the Cauchy function, Geman McClure function is selected as 2 which is selected experimentally based on the data.

*Evaluation metric*: The positioning performance of the listed six methods is evaluated in the east, north, and up (ENU) frame, based on the ground truth positioning provided by the NovAtel SPAN-CPT. As the GNSS positioning in the vertical direction is highly unreliable due to the satellite geometry, we mainly evaluate the horizontal positioning performance, including the mean error (MEAN), standard deviation (STD), maximum error (MAX), and the improvement of the new method compared with the FGO.

B. Experimental Evaluation in Urban Canyon 1

*1) Positioning Performance Comparison*

The positioning performance of the listed GNSS positioning methods is shown in Table 1. 17.39 meters of mean error is obtained using WLS with a standard deviation of 16.01 meters. After integrating the Doppler measurements using the EKF estimator, the error is reduced to 13.61 meters. Meanwhile, the maximum error reaches 88 meters due to the high-rising surrounding buildings leading to numerous GNSS outlier measurements. Fortunately, 9.45 meters of mean error is achieved using FGO with a standard deviation of 8.06 meters. However, the maximum error still reaches 31.94 meters. After applying the Cauchy function, the mean error decreases to 8.97 meters. The improvement shows that the Cauchy function can help to mitigate the effects of the outliers. However, the improvement is limited as the non-convex surface from Cauchy can introduce the local minimum issue. Meanwhile, the performance of the Cauchy function relies on the tuning of the shape parameters. Our previous work in [47] shows similar issues in visual positioning. Since the tuning of the shape parameters is out of the scope of this paper, we directly use the same shape parameters together with the Geman McClure function. After applying the Geman McClure function to the FGO (FGO-GM), the mean error decreases to 8.20 meters with a standard deviation of 6.61 meters. However, the improvement is still limited due to the non-convexity arising from the Geman McClure function. With the help of the GNC via FGO-GNC, the error decreases to 6.65 meters with a significantly reduced standard deviation of 4.81 meters, which shows the effectiveness of the proposed FGO-GNC. The last three rows in Table 1 show the 3D positioning performance using the listed six methods. A mean error of 20.32 meters is obtained using the FGO which is significantly larger than the horizontal error (9.45 meters in Table 1). With the help of the proposed method, the 3D mean error decreases to 14.72 meters. We can see that the improvement is also obtained in the vertical direction.

Figs. 7 and 8 show the 2D positioning error and derived trajectories, respectively. We can see that the positioning error is significantly reduced near epochs 350 and 390 using the FGO-GNC (blue curve). However, the error can still reach 20 meters even using the FGO-GNC. Different from the problem investigated in [32] where only the fully healthy measurements remain after de-weighting the outlier measurements, the majority of the GNSS measurements can be outliers although the magnitude of the error corresponding to each measurement can be different. Meanwhile, all the GNSS measurements involve atmosphere error. Therefore, it is difficult to fully identify outlier measurements among all the GNSS measurements in this case. However, the positioning accuracy is still improved by de-weighting the distinct outlier measurements.

As shown in Algorithm 1, the weightings of the pseudorange measurements are estimated gradually via iterations. To this end, it is interesting to see how the weightings of all the considered pseudorange measurements change during the iterations. The following Fig. 9 shows the estimated weightings for all the GNSS measurements under the conditions of the different maximum number of iterations. Fig. 9-(a) shows the weightings of the GNSS pseudorange measurements using FGO-GNC where we limit the maximum iterations of 1 in Algorithm 1. We can see from Fig. 9-(a) that only a small portion of the GNSS measurements are classified as outliers with smaller weightings (less than 1). With the increased number of iterations, the number of GNSS measurements with a weighting of less than 1 increased accordingly (e.g 5 iterations for Fig. 9-(b), 10 iterations for Fig. 9-(c), and 15 iterations for Fig. 9-(d)). Note that the FGO-GNC converges when the number of iterations reaches 15, satisfying the convergence condition of $\theta < 1$. We can see from Fig. 9-(d) that some of the weightings of the GNSS measurement are reduced to less than 0.1 which potentially denote the outlier measurements. Therefore, the effects of the outlier measurements are mitigated by the reduced weightings.

Table 1. Positioning performance of the listed six methods in urban canyon 1 (2D: horizontal positioning)

| All Data | WLS | EKF | FGO | FGO-Cauchy | FGO-GM | FGO-GNC |
|---|---|---|---|---|---|---|
| **2D MEAN (m)** | 17.39 | 13.61 | 9.45 | 8.97 | 8.20 | 6.65 |
| **2D STD (m)** | 16.01 | 15.19 | 8.06 | 7.58 | 6.61 | 4.81 |
| **2D Max (m)** | 94.43 | 88.97 | 31.94 | 36.48 | 34.11 | 24.09 |
| **2D Improvement** | | | | 5.07 % | 13.2 % | 29.63 % |
| **3D MEAN (m)** | 49.16 | 30.16 | 20.32 | 27.21 | 18.69 | 14.72 |
| **3D STD (m)** | 40.82 | 31.50 | 14.12 | 36.02 | 10.82 | 7.57 |
| **3D Max (m)** | 203.23 | 188.68 | 77.61 | 161.58 | 63.24 | 47.14 |
| **3D Improvement** | | | | -33.90% | 8.02% | 27.56% |

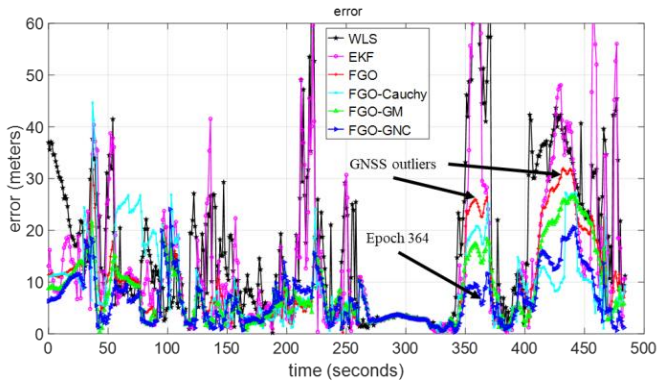

Fig. 7. 2D positioning errors of the evaluated six methods in urban canyon 1. The black and magenta curves denote the WLS and EKF, respectively. The red curve denotes the 2D error from FGO. The cyan, green and blue colors denote the FGO-Cauchy, FGO-GM, and FGO-GNC, respectively.

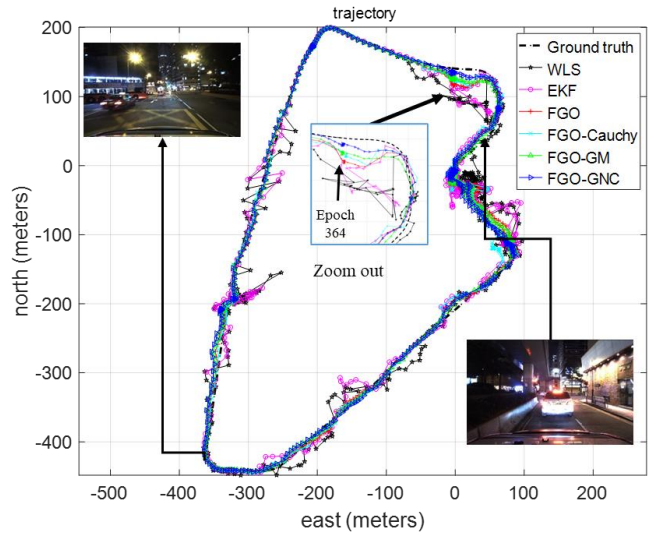

Fig. 8. 2D positioning trajectories of the evaluated six methods in urban canyon 1. The black dashed bold curve denotes the ground truth trajectory.

Fig. 10 shows the histogram of the GNSS pseudorange residuals using the FGO. We can see that the histogram involves the long tails on both the right and left sides. The long tail phenomenon is mainly caused by GNSS NLOS and multipath effects. According to our previous findings in [48], the noise of the pseudorange measurements in urban canyons is highly non-Gaussian. A Gaussian mixture model is an ideal option to describe the sensor noise of pseudorange measurements. Therefore, we parameterize the histogram by fitting the histogram using the GMM with three components as shown from the Table inside Fig. 10. The second component with a mean of -0.32 meters is estimated and the weighting reaches 0.23. Meanwhile, the third component with a mean of -3.8 meters makes up 27% of the potential GNSS outlier measurements. The histogram of the pseudorange residual derived from the FGO-GNC is shown in Fig. 11. Firstly, the weighting of the third component is significantly decreased from 0.27 to 0.09 which means that the residuals caused by the outlier measurements are mitigated. Secondly, the mean of the second component is also shifted closer to the zero means which is also caused by the mitigated residuals. Interestingly, we can see residuals corresponding to the outlier measurements that the long tail phenomenon still exists even using the FGO-GNC. By this observation, we believe that the combination of the FGO-GNC and the GMM is a promising solution to mitigate the effects of GNSS outliers complementarily, where the GMM can be used to model the remaining long-tail phenomenon after removing distinct outliers using the FGO-GNC.

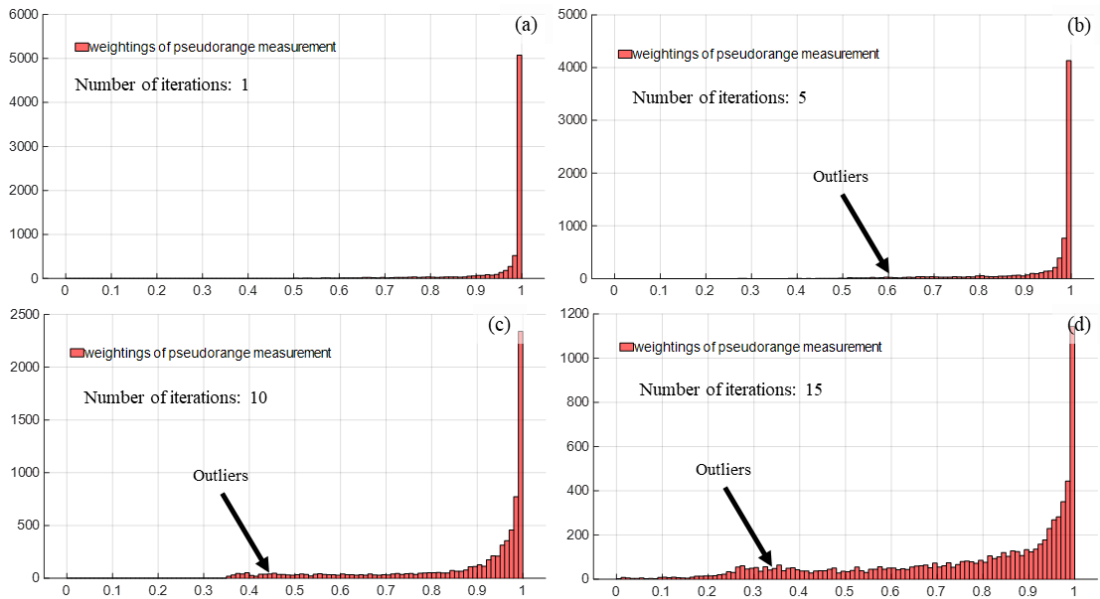

Fig. 9. Histogram of the weightings of all the measurements after applying the FGO-GNC in urban canyon 1. (a) FGO-GNC 1 iteration. (b) FGO-GNC with 5 iterations. (c) FGO-GNC with 10 iterations. (d) FGO-GNC with 15 iterations. The x-axis denotes the weightings ranging from 0 to 1. The y-axis represents the counts.

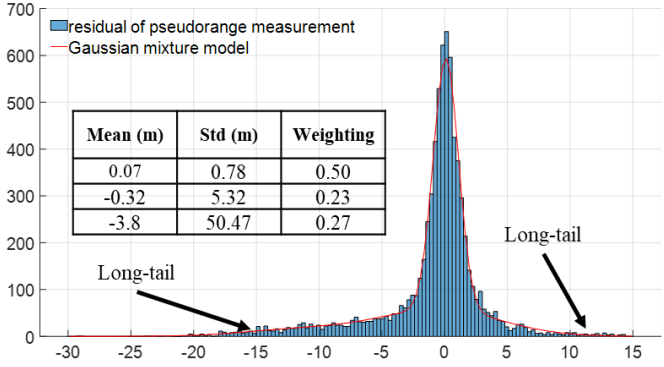

Fig. 10. Histogram of the GNSS pseudorange measurement residuals using the FGO. The axis denotes the residuals. The y-axis represents the counts.

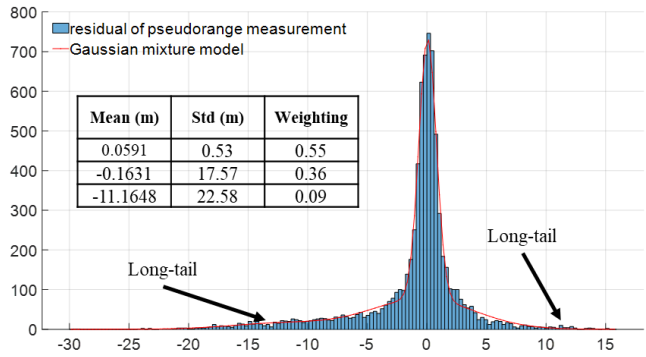

Fig. 11. Histogram of the GNSS pseudorange measurement residuals using the FGO-GNC. The axis denotes the residuals. The y-axis represents the counts.

Overall, the total number of iterations required by the proposed FGO-GNC is determined by the initial value of $\theta$ which is calculated by the (23). The larger $e_{r,t_{max}}^s$ requires more iterations. According to the evaluated dataset in an urban canyon, 16 iterations are enough for solving the proposed FGO-GNC. In short, the best positioning performance of GNSS pseudorange/Doppler integration is obtained using the proposed FGO-GNC. However, the remaining average error can still reach 6.65 meters.

*2) Case Study of the Coarse-to-Fine GNC process for FGO*

To show how the GNC helps to achieve the coarse-to-fine process, we present the positioning accuracy of the FGO-GNC at each iteration in Fig. 12. According to Section IV-B, the non-convexity of the Geman McClure function is recovered gradually with the decreased control parameter $\theta$. The x-axis in Fig. 12 denotes the number of iterations and the y-axis of the top panel denotes the mean error derived from FGO-GNC with a given number of iterations. We can see that the mean error (red curve in Fig. 12) decreases gradually with increased iterations. The bottom panel shows the value of the control parameter $\theta$ which reaches more than 100. Finally, the $\theta$ decreases to less than 1 and reaches the convergence criteria.

During the implementation, the scaling factor of $\theta$ during each iteration is set to 1.4 [33]. The smaller $\theta$ can introduce an smoother relaxation of the Geman McClure function from convex to non-convex. However, it can lead to high computation load. The larger $\theta$ can lead to faster convergence speed of the relaxation but leads to an unstable initial guess for the next iteration. Therefore, how to wisely select the scaling factor is still an open question.

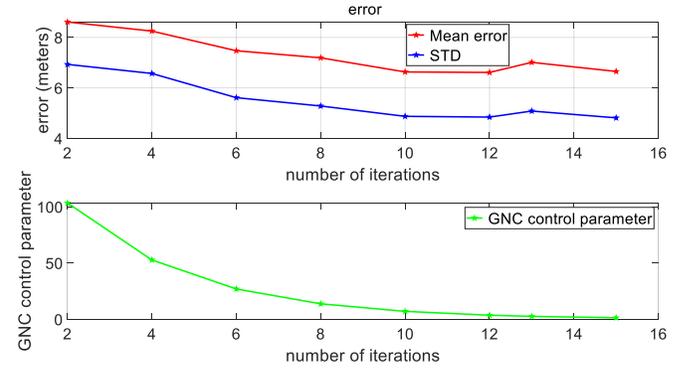

Fig. 12. Mean error and the standard deviation of the positioning of using FGO-GNC under different iterations. The x-axis denotes the number of iterations. The y-axis represents the error values for the top panel and the GNC control parameter ($\theta$) for the bottom panel.

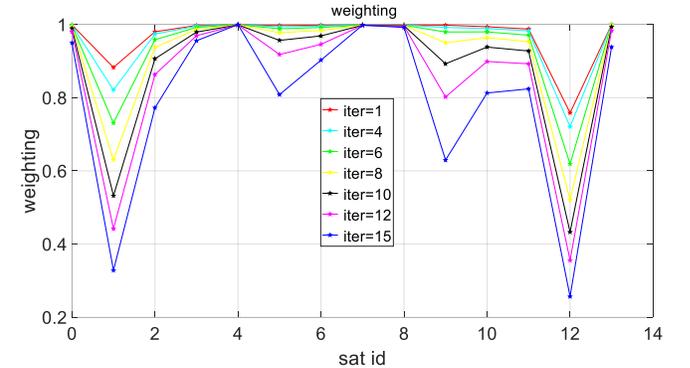

Fig. 13. Weightings of the pseudorange measurements at an epoch 47065 in different iterations of GNC. The x-axis denotes the id of the satellite (corresponding to the id of measurement). The y-axis denotes the value of the weightings.

The nature of the FGO-GNC in mitigating the effects of outlier measurements is to gradually estimate the weightings of the measurements via GNC. To show the detail of the weightings estimation during the GNC process, we select epoch 364 (see Fig. 7) as an example. Fig. 13 shows how the weightings for the 14 pseudorange measurements (GPS/BeiDou) change during the iterations. At the first iteration (see red curve), almost all the weightings are larger than 0.8. With the help of iterations of GNC, the difference between the inlier and outlier measurements becomes clear with some of the weightings being reducing to less than 0.4 (e.g. satellite 1, satellite 12 of the blue curve in Fig. 13). Meanwhile, some satellites maintain high weightings (around 1) even after the 15 iterations. Table 2 shows the mean error of the four different methods at epoch 47065. We can see that the FGO obtains the smallest 2D positioning error of 5.79 meters after deweighting the outlier measurements which shows the effectiveness of the proposed method.

Table 2. Positioning performance of the listed four methods at the epoch 364. (2D: horizontal positioning)

| Error (meter) | FGO | FGO-Cauchy | FGO-GM | FGO-GNC |
|---|---|---|---|---|
| 2D | 22.57 | 17.36 | 13.66 | 5.64 |
| 3D | 23.64 | 17.37 | 15.52 | 5.79 |

Table 3. Pseudorange errors of Epoch 47065 in the dataset collected the urban canyon 1

| Sat ID | Ele (degree) | C/$N_0$ (dB-Hz) | Type | Pseudorange Error (m) | $\omega_{t,s}$ |
|---|---|---|---|---|---|
| 0 | 51.26 | 19 | Multipath | 2.65 | 0.94 |
| 1 | 43.62 | 19 | NLOS | 61.99 | 0.32 |
| 2 | 32.58 | 28 | LOS | 6.1 | 0.77 |
| 3 | 41.13 | 39 | LOS | -1.5 | 0.95 |
| 4 | 59.61 | 39 | LOS | 0 | 0.99 |
| 5 | 50.59 | 34 | LOS | 0.84 | 0.80 |
| 6 | 64.33 | 34 | LOS | 0 | 0.90 |
| 7 | 32.89 | 31 | LOS | 0.66 | 0.99 |
| 8 | 48.37 | 30 | LOS | -0.37 | 0.99 |
| 9 | 48.61 | 14 | NLOS | 28.85 | 0.62 |
| 10 | 26.31 | 16 | NLOS | 8.57 | 0.81 |
| 11 | 39.30 | 36 | LOS | 0.1 | 0.82 |
| 12 | 29.92 | 24 | NLOS | 95.76 | 0.25 |
| 13 | 42.54 | 32 | NLOS | -1.65 | 0.93 |

To further verify the correctness of the weighting estimation, we present the ground truth pseudorange error and the LOS/NLOS/multipath classification results corresponding to the measurements at epoch 47065 in Table 3, using the double-difference technique of our team's previous work in [49]. The second and third columns denote the elevation angle and the carrier to noise ratio (C/$N_0$) of satellites. The fourth column represents the signal type classification results. The fifth column denotes the ground truth pseudorange error. Interestingly, satellite 1 introduces a pseudorange error of 61.99 meters due to the NLOS reflection. Meanwhile, the estimated weighting is less than 0.4 which can be seen in Fig. 13 using FGO-GNC. Similarly, the NLOS satellite 12 with a pseudorange error of 95.76 meters is also deweighted (weighting is less than 0.3). In short, the weightings are reduced corresponding to the multipath and NLOS signals. As a result, the 2D and 3D positioning accuracy is improved from 22.57 to 5.64 meters and 23.64 to 5.79 meters respectively which can be seen in Table 2. However, the proposed FGO-GNC can only mitigate the effects of the multipath/NLOS affected signals. The remaining average error can still reach more than 5 meters in the selected epoch 47065 which means that the upper bound of the proposed method relies on the number and quality of the other healthy GNSS measurements.

C. *Experimental Evaluation in Urban Canyon 2*

*1) Positioning Performance Comparison*

To challenge the performance of the FGO-GNC, we perform the other experiment in urban canyon 2 with denser buildings compared with that in urban canyon 1. In other words, urban canyon 2 involves a larger percentage of multipath/NLOS signals. In this experiment validation, we mainly evaluate the horizontal positioning performance of the listed six GNSS positioning methods which are shown in Table 4. 16.71 meters of mean error is obtained using WLS with a standard deviation of 16.01 meters. Due to the challenging building structures, the maximum 2D error reaches 112.15 meters. Similar to the phenomenon in the urban canyon 1 experiment, the improvement is also obtained via the EKF estimator. 12.03 meters of mean error is achieved using FGO, which is larger than the one in urban canyon 1 (9.45 meters) due to the denser buildings. After applying the Cauchy function, the mean error decreases to 10.02 meters which shows similar improvement. Interestingly, the mean error increases to 13.21 meters after applying the Geman McClure function, this is mainly caused by the local minimum issues arising from the non-convexity of the roust function. With the help of the GNC via FGO-GNC, the error decreases to 8.90 meters with a significantly reduced standard deviation of 6.28 meters, which shows the effectiveness of the GNC.

Figs. 14 and 15 show the 2D positioning errors and the trajectories of the listed four methods. Parts of the urban scene are shown in Fig. 15, where both sides are filled with tall buildings. Meanwhile, the high-rising double-decker bus can also lead to potential signal blockage and reflections. We can see that the error curve derived from the FGO-GNC (blue curve in Fig. 14) is significantly smoother compared with the original FGO (red curve in Fig. 14). This is caused by the mitigation of the outliers using the proposed method. The histograms of the weightings are shown in the appendix of this paper under different numbers of GNC iterations.

In short, the remaining positioning error still reaches 8.9 meters even applying the FGO-GNC. This is because, in some epochs, it is hard to get fully healthy GNSS measurements in such a dense urban canyon surrounded by tall buildings. In short, similar improvement is also achieved even in such a highly urbanized area.

Table 4. Positioning performance of the listed six methods in urban canyon 2 (2D: horizontal positioning)

| All Data | WLS | EKF | FGO | FGO-Cauchy | FGO-GM | FGO-GNC |
|---|---|---|---|---|---|---|
| 2D MEAN (m) | 16.71 | 13.20 | 12.03 | 10.02 | 13.21 | 8.90 |
| 2D STD (m) | 12.77 | 9.48 | 8.45 | 7.76 | 9.48 | 6.28 |
| 2D Max (m) | 112.15 | 52.17 | 43.01 | 51.24 | 52.17 | 41.10 |
| 2D Improvement | | | | 16.71 % | -9.8 % | 26.02 % |

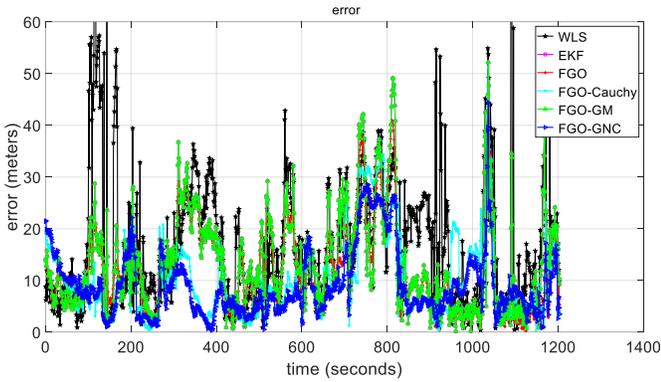

Fig. 14. 2D positioning errors of the evaluated six methods in urban canyon 2. The black and magenta curves denote the WLS and EKF, respectively. The red curve denotes the 2D error from FGO. The cyan, green and blue colors denote the FGO-Cauchy, FGO-GM, and FGO-GNC, respectively.

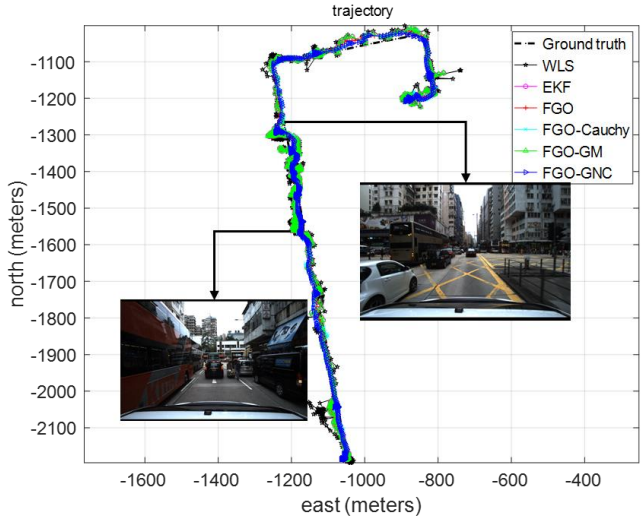

Fig. 15. 2D positioning trajectories of the evaluated six methods in urban canyon 2. The black and magenta curves denote the WLS and EKF, respectively. The red curve denotes the trajectory from FGO. The cyan, green and blue colors denote the trajectories from FGO-Cauchy, FGO-GM, and FGO-GNC, respectively. Meanwhile, the black dashed bold curve denotes the ground truth trajectory.

### D. Discussion: Positioning Performance Validation of Huawei P40 Pro Receiver

According to the experimental validation in urban canyons 1 and 2, the proposed FGO-GNC outperforms the conventional FGO method with significant improvements. However, the low-cost u-blox GNSS receiver belongs to the commercial level that is usually used in intelligent transportation (ITS) application and the patch antenna is the middle level. Therefore, we further investigate how the proposed FGO-GNC works for the low-cost mobile phone receiver. Different from the u-blox receiver, the internal antenna of the mobile phone is sensitive to multipath/NLOS signals and the collected GNSS measurements are noisy. The applied mobile phone is the latest Huawei P40 Pro (see Fig. 6) and raw GPS/BeiDou measurements are collected at a frequency of 1 Hz. The period of the data collection is about 1000 seconds.

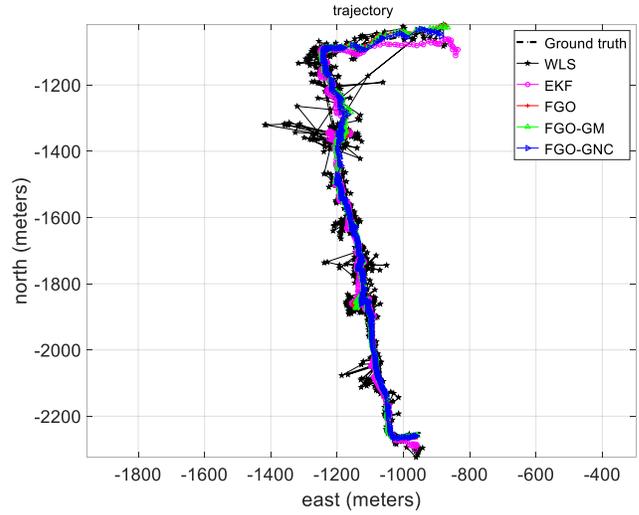

Fig. 16. 2D positioning trajectories of the evaluated six methods in urban canyon 2 using Huawei P40 Pro. The black and magenta curves denote the WLS and EKF, respectively. The red curve denotes the trajectory from FGO. The green and blue colors denote the trajectories from FGO-Cauchy, FGO-GM, and FGO-GNC, respectively. Meanwhile, the black dashed bold curve denotes the ground truth trajectory.

The positioning performance of the listed six methods is shown in Table 5. 28.0 meters of mean error is obtained using WLS with a standard deviation of 28.52 meters. Meanwhile, the maximum 2D error reaches 375.47 meters. The mean error decreases to 17.42 meters after applying the EKF estimator with a similar improvement compared with the u-blox GNSS receiver. The mean positioning error of FGO reaches 15.68 meters based on the GNSS measurements from the Huawei P40 Pro. The FGO-Cauchy fails to estimate the GNSS solutions due to the divergence. This is caused by the non-convex and truncated robust function of the Cauchy function. With the help of the proposed FGO-GNC, the mean error decreases to 13.6 meters with an improvement of 13.26%. The trajectories are shown in Fig. 16. To show the poor quality of the collected GNSS measurements from Huawei P40 Pro, the WLS results are denoted by the magenta color which deviates significantly from the ground truth trajectory. However, a significantly smoother and accurate trajectory is obtained using the proposed FGO-GNC (blue curve). In short, the proposed FGO-GNC leads to improved positioning accuracy even using the low-cost Huawei P40 Pro GNSS receiver.

Table 5. Positioning performance of the listed six methods in urban canyon 2 using Huawei P40 Pro (2D: horizontal positioning)

| **All Data** | **WLS** | **EKF** | **FGO** | **FGO-Cauchy** | **FGO-GM** | **FGO-GNC** |
|---|---|---|---|---|---|---|
| **2D MEAN (m)** | 28.00 | 17.42 | 15.68 | Fail | 15.09 | 13.6 |
| **2D STD (m)** | 28.52 | 12.31 | 9.87 | Fail | 9.34 | 7.38 |
| **2D Max (m)** | 375.47 | 69.22 | 52.97 | Fail | 58.94 | 48.62 |
| **2D Improvement** | | | | Fail | 3.76% | 13.26 % |

The elevation angle and the SNR are effective indicators of GNSS measurement quality. According to [50], the satellite signal with a lower elevation angle is more likely to be affected by surrounding environmental structures. Meanwhile, a signal with a smaller SNR value is likely to be attenuated by surroundings due to the signal reflection. Therefore, it is interesting to see how the filter using elevation angle and SNR threshold to remove the potential polluted GNSS signals. To this end, we heuristically set up several thresholds for both the elevation angle and SNR and the results are shown in Table 6. Interestingly, the error decreases from 15.68 meters (FGO in Table 5) to 15.01 meters after applying the elevation angle filtering with a threshold of 20°. However, the improvement is limited after increasing the threshold to 30°. Similarly, the improvement is also obtained after applying the SNR filtering with thresholds of 20. However, the positioning error even increases slightly after increasing the thresholds to 30. One of the reasons is that the exclusion of satellites can distort the satellite distribution and a similar phenomenon was also found in our previous work in [17, 51]. In short, the elevation angle and SNR thresholds based filtering can introduce improvement by excluding possible polluted GNSS signals. However, the improvement is limited by the percentage of the healthy GNSS measurements and the tuning of the thresholds.

Table 6. Positioning performance of the elevation/SNR filtering-based methods in urban canyon 2 using Huawei P40 Pro

| All Data | FGO (ele>20°) | FGO (ele>30°) | FGO (SNR>20) | FGO (SNR>30) |
|---|---|---|---|---|
| **MEAN (m)** | 15.01 | 14.97 | 15.17 | 15.85 |
| **STD (m)** | 9.40 | 8.46 | 9.57 | 9.53 |
| **Max (m)** | 52.69 | 52.35 | 46.29 | 45.00 |

*E. Discussion: Comparison of the Computational Time*

Different from the conventional filtering based, e.g. EKF, estimator, the proposed FGO-GNC employs the historical information into the optimization. Therefore, it is interesting to see the computational load caused using different methods. Table 7 shows the computational time used for the six methods. Both WLS and the EKF introduce low computational load with a mean time of less than 1 millisecond.

The computational time is significantly increased to 203.18 ms using FGO where all the historical information is considered. Slightly increased computational load is caused by the FGO-Cauchy and FGO-GM. Due to the multiple iterations required by the FGO-GNC, the computational time increased by triple reaching 711.65 ms. In short, the FGO-GNC significantly increases the computational load, compared with the conventional EKF estimator. One potential solution to decrease the computational load is to employ the sliding window technique [30] where only the historical information inside a sliding window is considered in FGO, which can be one of our future works.

Table 7. The computational time of the listed six methods in urban canyon 1 (millisecond: ms)

| Items | WLS | EKF | FGO | FGO-Cauchy | FGO-GM | FGO-GNC |
|---|---|---|---|---|---|---|
| **MEAN (ms)** | 0.55 | 0.47 | 203.18 | 227.84 | 250.72 | 711.65 |
| **STD (ms)** | 0.20 | 0.17 | 12.20 | 28.46 | 32.18 | 20.15 |
| **Max (ms)** | 0.61 | 0.52 | 215.73 | 240.17 | 263.59 | 735.71 |

VI. CONCLUSIONS AND FUTURE WORK

GNSS positioning is currently still the major source of globally referenced positioning for intelligent transportation systems (ITS). However, accurate GNSS positioning in urban canyons is still a challenging problem. Multipath/NLOS phenomenons currently remain the major problems for GNSS positioning in urban canyons. This paper proposed an FGO-GNC to improve the GNSS positioning accuracy by mitigating the effects of potential outlier measurements. The effectiveness is verified by several datasets collected in urban canyons of Hong Kong, via both the ITS-level and mobile phone level GNSS receivers. Meanwhile, the proposed method does not require additional sensors.

In the future, we will apply the proposed FGO-GNC to carrier-phase measurements for GNSS real-time kinematic positioning. With the increased availability of multiple satellite constellations, more and more satellites are visible to GNSS receivers. It is interesting to investigate the effectiveness of the proposed method in multiple satellite constellations and frequencies. As evaluated using the Huawei P40 Pro receiver, the remaining error of the FGO-GNC can still reach 13.6 meters due to the poor quality of the GNSS measurements. However, the dynamic model of the vehicle, inertial measurement unit (IMU) inside the phone are promising sources to improve the overall positioning accuracy which will also be part of our future work.

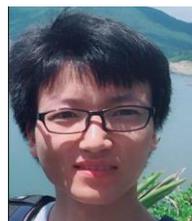

**Weisong Wen** was born in Ganzhou, Jiangxi, China. He received a Ph.D. degree in mechanical engineering, the Hong Kong Polytechnic University, in 2020. He is currently a senior research fellow at the Hong Kong Polytechnic University. His research interests include multi-sensor integrated localization for autonomous vehicles, SLAM, and GNSS positioning in urban canyons. He was a visiting student researcher at the University of California, Berkeley (UCB) in 2018.

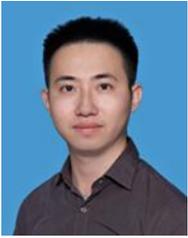

**Guohao Zhang** received a bachelor's degree in mechanical engineering and automation from the University of Science and Technology Beijing, China, in 2015. He received a master's degree in Mechanical Engineering and currently pursuing a Ph.D. degree in the Interdisciplinary Division of Aeronautical and Aviation Engineering, the Hong Kong Polytechnic University. His research interests including GNSS urban localization, cooperative positioning, and multi-sensor integrated navigation.

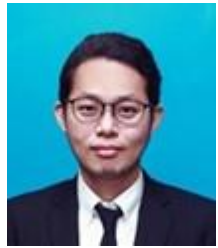

**Li-Ta Hsu** received the B.S. and Ph.D. degrees in aeronautics and astronautics from National Cheng Kung University, Taiwan, in 2007 and 2013, respectively. He is currently an assistant professor with the Interdisciplinary Division of Aeronautical and Aviation Engineering, The Hong Kong Polytechnic University, before he served as a post-doctoral researcher in the Institute of Industrial Science at the University of Tokyo, Japan. In 2012, he was a visiting scholar at University College London, the U.K. His research interests include GNSS positioning in challenging environments and localization for pedestrian, autonomous driving vehicle, and unmanned aerial vehicle.

APPENDIX

The histograms of the weightings corresponding to the experimental validation in urban canyon 2 are shown in the following Fig. 17 under different numbers of GNC iterations.

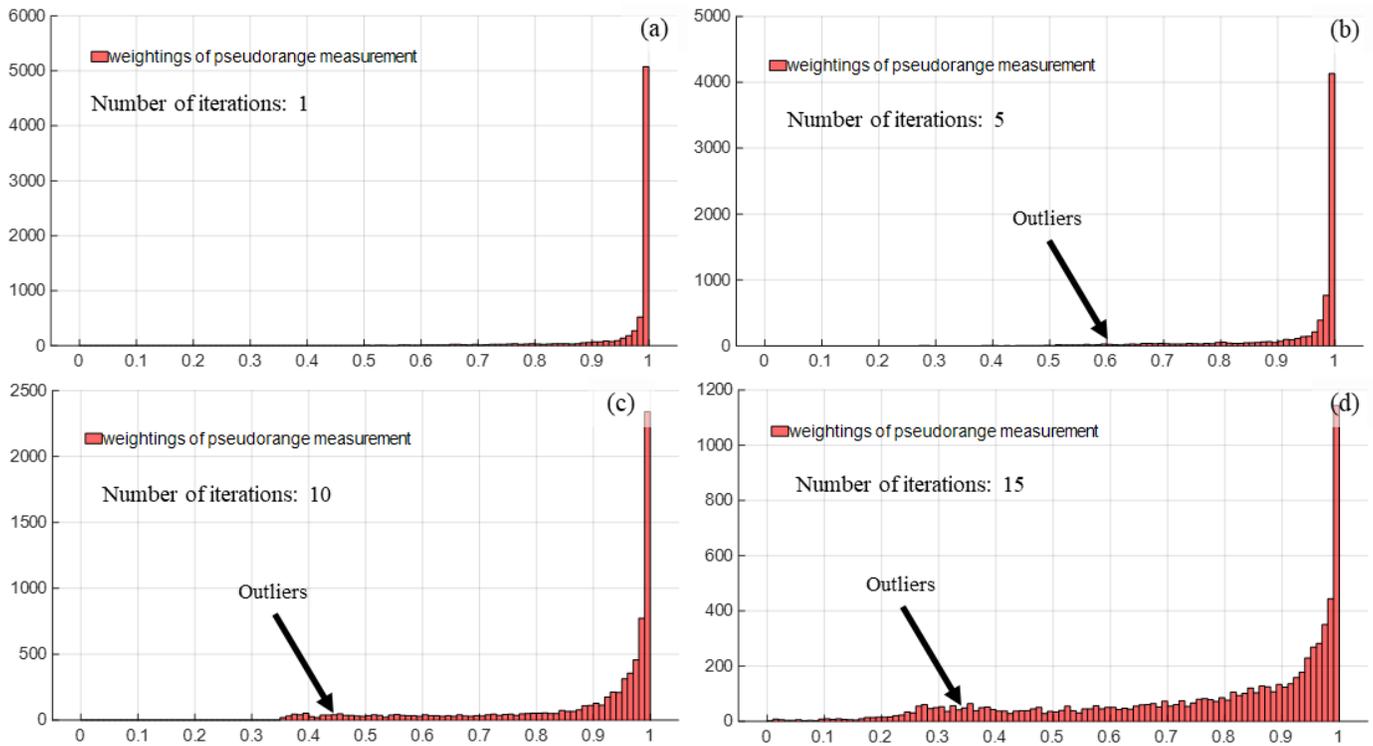

Fig. 17. Histogram of the weightings of all the measurements after applying the FGO-GNC in urban canyon 2. (a) FGO-GNC 1 iteration. (b) FGO-GNC with 5 iterations. (c) FGO-GNC with 10 iterations. (d) FGO-GNC with 15 iterations. The x-axis denotes the weightings ranging from 0 to 1. The y-axis represents the counts.